\documentstyle[12pt]{article}
\begin{document}
\input epsf
\title{Mesoscopic Fluctuations in Stochastic Spacetime}
\author{K. Shiokawa\thanks
{  E-mail address: kshiok@phys.ualberta.ca
 }
\\
{\small Theoretical Physics Institute}\\ {\small University of
Alberta, Edmonton, Alberta T6G 2J1, Canada}\\ 
\small{\it(Alberta Thy 23-99)} }
\maketitle
\begin{abstract}
Mesoscopic effects associated with wave propagation in spacetime
with metric stochasticity are studied. We show that the scalar and
spinor waves in a stochastic spacetime behave similarly to the
electrons in a disordered system. Viewing this as the quantum
transport problem, mesoscopic fluctuations in such a spacetime are
discussed. The conductance and its fluctuations are expressed in
terms of a nonlinear sigma model in the closed time path
formalism. We show that the conductance fluctuations are
universal, independent of the volume of the stochastic region and
the amount of stochasticity.
\end{abstract}
%
{\bf To appear in Physical Review D}
\newpage
\section{Introduction}
\subsection{Metric fluctuations in semiclassical gravity}

Since Einstein-Hilbert action is insensitive to fluctuations
near the Planck scale, we expect large fluctuations in this regime
to occur which may require the reconsideration of the concept of
microstructure of spacetime itself. The concept of metric
fluctuations thus introduced originally by Wheeler
\cite{Wheeler57,MTW} have been studied and modified in various
different contexts. Spacetime foams or other string theory
motivated microstructure of spacetime can be treated as possible
stochastic sources \cite{Foam}.
 Since those fluctuations become dominant only near the Planck scale and cannot be
directly observable, it is important to identify their possible
influence on the matter fields propagating in such a background
with the energy much lower than the Planck value which is possibly
detectable by high energy astrophysical observation.

Since quantum gravity is still an unsolved problem, in the
conventional semiclassical gravity, a spacetime is left
unquantized \cite{BirrellDavis82}.
 The recent study in semiclassical gravity, however, reveals
 many pathological features in this approach.
One of the most serious problems is the violation of positive
energy theorem due to large quantum fluctuations which leads to
the violation of causality by allowing the creation of traversable
wormholes \cite{KuoFord93}. It is also noticed that the positivity
can be recovered by imposing the additional smearing in the case
of Minkowski spacetime \cite{FlanaganWald96}. This smearing,
originated in the microscopic quantum fluctuations, may also cure
other problems such as initial or blackhole singularity problems.
Other studies also suggest that the small stochastic fluctuations
of spacetime metric lying on the deterministic background spactime
are not only the useful phenomenological modification of
semiclassical Einstein equation but also the inevitable
consequence of the more fundamental processes or of the
backreaction induced by matter fluctuations
\cite{HuMatacz95,CamVer96}. In an astrophysical context, squeezed
states, evolved from primordial gravitational waves by parametric
amplification during the cosmological expansion \cite{Squeeze},
are considered to be one of the origins of the stochastic
gravitational waves believed to exist in the present universe.
Whether or not the trace of such primodial fluctuations as the
quatum noise lies within the detectable range of the Laser
Interferometric Gravitational Wave Observatory detector
(LIGO)\cite{LIGO92} is the curious question under debate
\cite{Grishchuck98,AFA99}.

\subsection{Semiclassical gravity and mesoscopic physics}
Semiclassical gravity, though far different in energy scale,
shares many common features with mesoscopic physics
\cite{Drexel,Hu99}. The quantum transport properties of metallic
systems are known to be divided into several different regimes
depending on the qualitatively different contribution of
scatterers.
For the length scale less than the mean free path $l_{M}$, the
wave propagates ballistically similar to the free coherent wave;
for the length scale larger than the coherence length $L_{coh}$,
the scale at which the mutual coherence of waves is lost due to
inelastic scattering, the classical Boltzmann transport theory is
valid. In the mesoscopic scale $l_{M} < L < L_{coh}$, multiple
scattering has to be taken into account and the coherence between
different paths becomes important. The effect of the environment
has to be taken into account and the dissipation and decoherence
due to inelastic scattering play an important role. Furthermore,
in the mesoscopic regime close to $L_{coh}$, the quantum-classical
correspondence of the propagating wave becomes relevant;
 in the same regime close to  $l_{M}$,
 possible influence of the microscopic
constituent of a medium will manifest itself on the transport
properties of the wave. Owing to the recent progress in nanoscale
technology \cite{Imry97}, many phenomena in this regime are
amenable to experiments. In light of this analogy, various effects
associated with the electromagnetic wave propagation in the
Friedmann-Robertson-Walker universe and the Schwarzschild
spacetime with metric stochasticity were studied in
\cite{HuShiokawa98} based on the formal equivalence of the
Maxwell's equations in a stochastic spacetime with those in random
media. Localization of photon and anomalous particle creation can
occur in such stochastic spacetimes.

In this paper, we show that the scalar and spinor fields
propagating in a stochastic Minkowski spacetime can be considered
as the electrons propagating in a disordered potential. The
randomness couples to the frequency of the wave additively for the conformal
metric fluctuations.
Thus the effect of the stochasticity
resembles that of a random potential.
We use the closed time path formalism that allows us to study
equilibrium and nonequilibrium quantum field theory in the unified
framework \cite{CTP1,CTP2}. The closed time path partition
function is defined for the stochastic quantum system from which
we obtain the effective interaction between fundamental fields in
the form of a four point vertex after averaging over
stochasticity. Diagonalization of the matrix Green function is
employed. This is an essential step to make use of the nonlinear
sigma model developed for the disordered transport problem
\cite{Wegner79,ELK80,Hikami81,PruSha82,BelKir94}. We use the
Hubbard-Storatonovich transformation and write the collective
excitations in terms of auxiliary fields. The conductance and its
fluctuations are expressed by these auxiliary fields.


The paper is organized as follows: In Sec. 2, we start from showing
that the effects of conformal metric fluctuations on scalar and
Dirac fields can be
simply represented by the effects of a fluctuating mass.
 In Sec. 3, we develop the nonlinear
sigma model to express higher order fluctuations of fundamental
fields in terms of the correlation function of collective fields.
The conductance associated with propagation of particles in such a
spacetime can be defined analogous to the one in electric circuits
by the Kubo formula as the correlation function of currents.
Similar to the mesoscopic quantum transport problem, we show that
the conductance fluctuations are universal, independent of the
size of the stochastic region and the amount of stochasticity
initially assumed. The amplitude of the conductance fluctuations
is constant upto the leading order in the weak disorder expansion.
In Sec. 4, the summary of results is given followed by brief
discussions.

\section{Wave Propagation in Stochastic Spacetimes}

 \subsection{Scalar and Dirac fields in stochastic spacetimes}

The Lagrangian density for the complex free scalar field in curved spacetime
has the form
\begin{eqnarray}
\label{Lagrangian}
  {\cal L_{S}} = \sqrt{-g} [
   g_{\mu \nu} \partial^{\mu} \phi^{\dag} \partial^{\nu} \phi
   - ( m_{S}^2 + \xi R ) \phi^2 ],
\end{eqnarray}
where $\xi$ is a dimensionless nonminimal coupling parameter.
$\xi= 1/6$ is called conformal coupling and $\xi = 0$ is minimal
coupling. In the presence of a slight amount of inhomogeneity in a
flat spacetime background characterized by $g_{\mu \nu} =
\eta_{\mu \nu} + h_{\mu \nu}$, the Lagrangian given above can be
split into the flat space term and the perturbation term as ${\cal
L}_S = {\cal L}_{S0} + {\cal L}_{SI}$:
\begin{eqnarray} \label{Lagrangian-flat}
  {\cal L}_{S0} =
   \eta^{\mu \nu} \partial_{\mu} \phi^{\dag} \partial_{\nu} \phi
   - m_{S}^2 \phi^2
\end{eqnarray}
and ${\cal L}_{SI} = - T_{S}^{\mu \nu} h_{\mu \nu}$, where
$T_{S}^{\mu \nu}$ is the stress-energy tensor. In a conformal
coupling case ($\xi = 1 / 6$), the action Eq. (\ref{Lagrangian})
is conformally invariant except for the mass term. Thus, a
conformal type of metric fluctuations can be attributed to the
effect of a fluctuating mass. Accordingly if we write the
stress-energy tensor in the following form
\begin{eqnarray} \label{Tab}
  T^{\mu \nu}_S &=& \partial^{\mu} \phi^{\dag} \partial^{\nu} \phi
  - \frac{1}{2} \eta^{\mu \nu} [
   \eta^{\lambda \rho} \partial_{\lambda} \phi^{\dag} \partial_{\rho} \phi
   + m_{S}^2 \phi^2 ]
   - 2 \xi [ \partial^{\mu} \phi^{\dag} \partial^{\nu} \phi
   - \eta^{\mu \nu} \eta^{\lambda \rho} \partial_{\lambda} \phi^{\dag}
    \partial_{\rho} \phi
   \nonumber  \\
   &+& \phi^{\dag} \partial^{\mu} \partial^{\nu} \phi
   - \frac{1}{4} \eta^{\mu \nu} \phi^{\dag} \Box \phi
   - \frac{3}{4} \eta^{\mu \nu} m_{S}^2 \phi^2
   ]
%
%
\end{eqnarray}
for an isotropic and conformal type of stochasticity $g_{\mu \nu}
= \eta_{\mu \nu} e^{ v(x) } \sim \eta_{\mu \nu} + \eta_{\mu \nu}
v(x)$, where $v(x)$ is a stochastic field, the interaction term
can be simply written as
\begin{eqnarray} \label{flucmass}
 {\cal L}_{SI} = - m_{S}^2 v(x) \phi^2(x).
\end{eqnarray}
In this case the total Hamiltonian is given by
\begin{eqnarray} \label{RealHamiltonian}
  H &=&
   \int d^3 x
     [ \pi^2 + (\nabla \phi)^2 + m_{S}^2 \phi^2
     + m_{S}^2 v(x) \phi^2 ],
\end{eqnarray}
or in a momentum representation,
\begin{eqnarray} \label{MomHamiltonian}
  H &=&
     \sum_{p}
     [ \phi(-p) (p^2 + m_{S}^2) \phi(p)
           + m_{S}^2 \sum_{q} v(q)
     \phi(p) \phi(p + q)
     ].
\end{eqnarray}
Furthermore, if we restrict our case to elastic scattering, we
have
\begin{eqnarray} \label{NRHamiltonian}
  H &=&
     \sum_{\vec{p}}
     [ \phi(-\vec{p}) (\vec{p}^2 + m_{S}^2 - \omega^2) \phi(\vec{p})
     + m_{S}^2 \sum_{\vec{q}} v(\vec{q})
     \phi(\vec{p}) \phi(\vec{p} + \vec{q})
     ],
\end{eqnarray}
where $\omega \equiv p^0$. This has the form equivalent to
electron propagation in a disordered potential.

The equation of motion in this model will be
\begin{eqnarray}  \label{Start2}
\nabla^2 \phi + (\omega^2 - m_{S}^2) \phi + v_{S}(x) \phi
 = 0,
\end{eqnarray}
where $v_{S}(x) \equiv m_{S}^2 v(x)$. Comparing Eq. (\ref{Start2})
with the equation of motion for the electromagnetic wave
\cite{HuShiokawa98}
\begin{eqnarray}  \label{Start3}
\nabla^2 \phi + \omega^2 \phi - \omega^2 v_{S}(x) \phi = 0,
\end{eqnarray}
we see that the randomness couples additively to the frequency in
the scalar wave, whereas it couples multiplicatively in the
electromagnetic wave. Since the electromagnetic wave obeys
Maxwell's equations which are conformally-invariant, the equation
of motion is rather similar to that of wave propagation in random
media. While for the scalar wave subjected to conformal metric
fluctuations, the equation of motion has the form of massive
particles in a random potential. We expect that the property of 
fields under the influence of more general types of metric
fluctuations will possess both aspects.
The difference of the form between Eq. (\ref{Start2}) and Eq. (\ref{Start3})
leads to the different density of states and transport properties.
We also point out that although the restriction of the metric fluctuations
to the conformal type simplifies the arguments significantly,
many arguments in the rest of the paper is applicable to
 the more general class of fluctuations.


The Lagrangian density for the Dirac field in curved spacetime
has the form
\begin{eqnarray} \label{DiracLagrangian}
  {\cal L}_D = b \left[
  \frac{i}{2} \bar{\psi}
             \not{\!\nabla}
   \psi - m_D \bar{\psi} \psi
  \right],
\end{eqnarray}
where $  \not{\!\nabla} \equiv \gamma_{a} \nabla^{a}$, $
\gamma^{a} \equiv b^{a}_{~\mu} \gamma^{\mu}$, and $b = \det
b^{\mu}_{~a}$ for vierbein fields $b^{\mu}_{~a}$. The Lagrangian
given above splits into the flat space contribution and the
perturbation term as ${\cal L} = {\cal L}_{D0} + {\cal L}_{DI}$
similar to the scalar field case:
\begin{eqnarray} \label{DiracLagrangian-flat}
  {\cal L}_{D0} =
  \frac{i}{2} \bar{\psi}
             \not{\!\nabla}
   \psi - m_D \bar{\psi} \psi
\end{eqnarray}
and ${\cal L}_{DI} = - T^{\mu \nu}_{D} h_{\mu \nu}$, where $T^{\mu
\nu}_{D}$ is the stress energy tensor that can be written as
\begin{eqnarray} \label{DiracTab}
  T^{\mu \nu}_{D} &=&
   \frac{i}{2} \left[
      \bar{\psi} \gamma^{\mu} \nabla^{\nu} \psi
   - \nabla^{\nu} \bar{\psi} \gamma^{\mu}  \psi
   \right].
%
%
\end{eqnarray}
 For a massless Dirac field,
the action Eq. (\ref{DiracLagrangian}) is conformally invariant.
Thus, a conformal type of metric fluctuations can also be
considered as the effect of fluctuating mass as in the scalar
field. For an isotropic, conformal type of stochasticity $v(x)$
defined above, the interaction term can be manifestly given as
\begin{eqnarray} \label{Diracflucmass}
 {\cal L}_{DI} =   \frac{i}{2} v(x) \bar{\psi}(x)
             \not{\!\nabla}  \psi(x)
         = m_D v(x) \bar{\psi}(x) \psi(x),
\end{eqnarray}
where we used the equation of motion to obtain the second expression.
The equation of motion in this model will be
\begin{eqnarray}  \label{DiracStart2}
 \left(
 i \not{\!\partial} - m_D - v_{D}(x)
 \right)
 \psi(x) = 0,
\end{eqnarray}
where $v_{D}(x) \equiv m_D v(x)$.
 The corresponding equations for Green
functions are
\begin{eqnarray}
-( \Box + m_{S}^2 + v_{S}(x) ) G(x,x')
 = \delta^{4}(x - x') \mbox{ for a scalar field},
 \\
 \left(
 i \not{\!\partial} - m_D - v_{D}(x)
 \right) S(x,x')
 = \delta^{4}(x - x') \mbox{ for a Dirac field}.
\end{eqnarray}
%


 We expect that the autocorrelation
functions of the stochastic fields have the following forms:
\begin{eqnarray} \label{Delta}
 \Delta_{S}(x-y) &=& \langle v_{S}(x) v_{S}(y) \rangle =
 u_{S} {\cal F}_L( |\vec{x} - \vec{x'}| / L)
     {\cal F}_T( |t - t'| / T),
\nonumber  \\      \Delta_{D}(x-y) &=& \langle v_{D}(x) v_{D}(y) \rangle =
 u_{D} {\cal F}_L( |\vec{x} - \vec{x'}| / L)
     {\cal F}_T( |t - t'| / T),
%
%
%
\end{eqnarray}
%
where ${\cal F}_L$ and ${\cal F}_T$ are rapidly decaying functions
and $L$ and $T$ are characteristic correlation ranges of the
stochastic field in space and time, respectively. 
$L$ and $T$ can be regarded as minimum length and time 
appeared as a result of coarse graining of microscopic dynamics
characterizing the finite resolution of spacetime.
The functions
${\cal F}_L$ and ${\cal F}_T$ are normalized such that $\int {\cal
F}_L d^3 x = \int {\cal F}_T dt =1$. For a possible choice of the
form of fluctuations,
\begin{eqnarray} \label{Deltaex}
\langle v(x) v(y) \rangle =  \frac{ l_{pl}^{~\alpha} }{ |\vec{x} -
\vec{x'}|^{\alpha} } e^{-  |\vec{x} - \vec{x'}| / l_{pl} }
\end{eqnarray} \label{ufromml}
for a constant $\alpha$, where $L = l_{pl} = 1.6 \times 10^{-33}$
cm is the Planck length, we have
\begin{eqnarray}  \label{ufroml}
u_{S} = 4 \pi m_{S}^4 l_{pl}^3 ,
\end{eqnarray}
 independent of $\alpha$.
 The disorder-averaged retarded (advanced) Green function
 $\langle G_{R (A)}(p,p')\rangle = G_{R (A)}(p)$
 can be written in the form
\begin{eqnarray}  \label{Green}
G_{R (A)}(p)  = \frac{1}{p^2 -m_{R}^2 \mp i \mbox{~sign}(p_{0})
\Sigma_{I}(p)},
\end{eqnarray}
where $m_{R}$ is a renormalized mass and 
$\Sigma_{I}(p)$ is an imaginary part of the self energy. The
upper (lower) sign corresponds to the retarded (advanced) Green
function. 
%
For the stochastic field that obeys Eq. (\ref{Deltaex}),
$v(x)$ can be approximated as a white noise potential if
$|\vec{p}|<< M_{pl}$, where $M_{pl}= 10^{19} $ GeV is a Planck mass.
Consequently, the self energy $\Sigma_{I}(p)$ is independent of
the momentum and depends only on the frequency in such a limit.
That is, the low energy elastic scattering is
insensitive to the length scale that characterizes
the fine structure of the
medium. 
In this limit, the effect of the medium is to renormalize
the frequency of the wave propagating through it.  The real part
of the self energy can be absorbed in the frequency term in this
limit and will not be considered in this paper. Such a
medium is called the effective medium.
%
%
Under this condition, 
the mean free path of this system is given by $l_{M} =
4 \pi / u_{S}$ in the Born approximation. Combining with Eq.
(\ref{ufroml}), we obtain $l_{M} = m_{S}^{-4} l_{pl}^{-3}$. 
For a mass $m_{S}$ much smaller than the Planck mass ( $m_{S}
<< M_{pl}$), this naive calculation illustrates that 
the mean free path is much larger than the Planck
length ($l_{M} >> l_{pl}$). 
Similar results hold for the spinor field as well.
The coherence length $L_{coh}$, on the other hand, is determined
by many factors including the other possible interactions, the
time dependence in the randomness, and external fields and will
not be specified here.

\section{Mesoscopic Effects in Stochastic Minkowski Spacetime}
In this section, we consider quantization of the noninteracting
scalar field that obeys the stochastic equation of motion
discussed in the last section. The parallel argument for the Dirac
field is given in Appendix A. The stochastic action for the scalar
field has the form
\begin{eqnarray} \label{NLSBActionBar}
 S_{v_{S}}[\phi,\phi^{\dag}]=
 \int d^{4}x
\left[  \partial^{\mu} \phi^{\dag} \partial_{\mu} \phi
      -  m_{S}^2 \phi^2 - v_{S} \phi^2 \right].
\end{eqnarray}
This action is invariant under the global $U(1)$ gauge
transformation
\begin{eqnarray}       \label{NLSBglobalgauge}
      \phi(x)  &\rightarrow& e^{i \theta}   \phi(x), \nonumber  \\
 \phi^{\dag}(x) &\rightarrow& \phi^{\dag}(x) e^{-i \theta}.
\end{eqnarray}
The corresponding Noether current is
\begin{eqnarray}      \label{NLSB302}
         J_{\mu}(x) =
      i \phi^{\dag}  \partial_{\mu} \phi
- i  \partial_{\mu} \phi^{\dag}  \phi.
\end{eqnarray}

\subsection{Closed time path formalism for the stochastic system}

\begin{figure}[h]
  \begin{center}
\epsfxsize=.9\textwidth \epsfbox{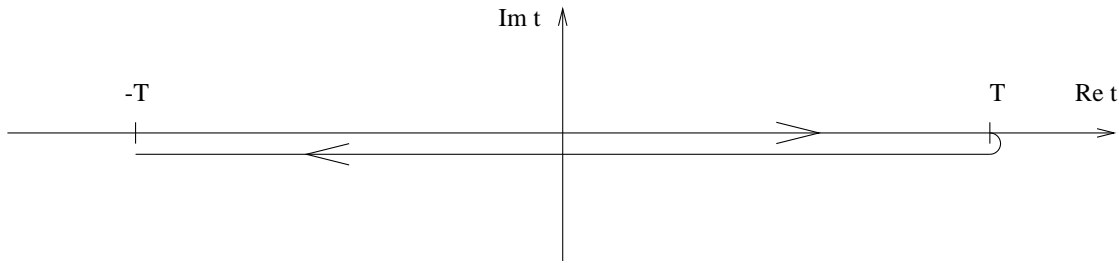}
  \end{center}
\caption{The contour in the closed time path
integral formalism.}
\label{fig1}
\end{figure}
The quantization of the stochastic system described in Eq.
(\ref{Start3}) can be treated in the closed time path formalism.
Here we assume that the whole system consists of a complex scalar
quantum field $\phi$ and a classical stochastic field $v_{S}$.
%
%
Then the density matrix for the whole system is given by
\begin{eqnarray}
\rho_{S}[\phi_{1},\phi_{1}^{\dag},\phi_2,\phi_{2}^{\dag},v_{S},t] =
 \langle \phi_{1},\phi_{1}^{\dag},v_{S} | \hat{\rho_{S}}(t)
 | \phi_2,\phi_{2}^{\dag},v_{S} \rangle,
\end{eqnarray}
where $| \phi,\phi^{\dag}, v_{S} \rangle $ is an eigenstate of the
field operator $\hat{\phi}$ for a particular realization of $v_{S}$ such that
\begin{eqnarray}
    \hat{\phi}(x) | \phi , \phi^{\dag},v_{S} \rangle &=&
          \phi(x) | \phi , \phi^{\dag},v_{S} \rangle.
\end{eqnarray}
To obtain the reduced density matrix
$\rho_{S}[\phi_{1},\phi_{1}^{\dag},\phi_2,\phi_{2}^{\dag},t]$
for the system, we average over $v_{S}$ as
\begin{eqnarray}
     \rho_{S}[\phi_{1},\phi_{1}^{\dag},\phi_2,\phi_{2}^{\dag},t] =
     \langle \rho_{S}[\phi_{1},\phi_{1}^{\dag},
     \phi_2,\phi_{2}^{\dag},v_{S},t] \rangle_v.
\end{eqnarray}

The closed time path partition function for this system
is given by
\begin{eqnarray}
 Z[J,J^{\dag},v_{S}]
   &=&  \int d\phi_{f} d\phi_{f}^{\dag}
   \langle 0_{-} | \tilde{T}
      \exp \left[  -i J_2 \cdot \hat{\phi}  \right]
    | \phi_{f},\phi_{f}^{\dag}, v_{S} \rangle
\nonumber  \\
    &\times&
      \langle \phi_{f},\phi_{f}^{\dag}, v_{S} | T
      \exp \left[  i J_1 \cdot \hat{\phi}
             \right]
    | 0_{-} \rangle
 \nonumber  \\
    &=&
    \int d\phi_{f} d\phi_{f}^{\dag}
    D\phi_1 D\phi_{1}^{\dag} D\phi_2 D\phi_{2}^{\dag}
      \exp \left[  i( S[\phi_1,\phi_{1}^{\dag} ] - S[\phi_2,\phi_{2}^{\dag} ]
      + J_1 \cdot \phi_1 - J_2 \cdot \phi_2 )
           \right]
     \nonumber  \\
    & &
    \times
   \exp \left[ i( S_I[\phi_1,\phi_{1}^{\dag},v_{S}] - S_I[\phi_2,\phi_{2}^{\dag},v_{S}] )  \right]
\end{eqnarray}
for  $\hat{\rho_{S}}(t_{i}) = | 0_{-} \rangle \langle  0_{-} |$
and $S_I[\phi,\phi^{\dag},v_{S}] \equiv - \int d^4 x ~v_{S}
\phi^2$, $J \cdot \phi \equiv \int d^4 x \left[ J^{\dag}(x)
\phi(x) + \phi^{\dag}(x) J(x) \right]$ and $\tilde{T}$ denotes an
anti-time-ordered product. The path integral above is defined
along the two paths, one forward in time and the other backward in
time (Figure 1). For an arbitrary initial state given by
$\rho_{S}[\phi_{1i},\phi_{1i}^{\dag},\phi_{2i},\phi_{2i}^{\dag}]$,
the partition function takes the form
\begin{eqnarray}
     \label{StocPart}
 Z[J,J^{\dag},v_{S}]
   &=&
    \int d\phi_{f} d\phi_{f}^{\dag}
    D\phi_1 D\phi_{1}^{\dag} D\phi_2 D\phi_{2}^{\dag}
     \exp \left[
       i( S[\phi_1,\phi_{1}^{\dag} ] - S[\phi_2,\phi_{2}^{\dag} ]
     + J_1 \cdot \phi_1 - J_2 \cdot \phi_2 )
          \right]
\nonumber  \\
   &\times&
    \rho_{S}[\phi_{1i},\phi_{1i}^{\dag},\phi_{2i},\phi_{2i}^{\dag}]
     F[\phi_{1},\phi_{1}^{\dag},\phi_2,\phi_{2}^{\dag}, v_{S}],
\end{eqnarray}
where
\begin{eqnarray}
    F[\phi_{1},\phi_{1}^{\dag},\phi_2,\phi_{2}^{\dag}, v_{S}]
    = \exp \left[
    i( S_I[\phi_1,\phi_{1}^{\dag},v_{S}] - S_I[\phi_2,\phi_{2}^{\dag},v_{S}])
           \right]
\end{eqnarray}
is a stochastic influence functional.

In the absence of the stochastic field,
\begin{eqnarray}
 \label{CTPPF}   
Z[J,J^{\dag}]
   &=&
    \int d\phi_{f} d\phi_{f}^{\dag}
    D\phi_1 D\phi_{1}^{\dag} D\phi_2 D\phi_{2}^{\dag}
     \exp \left[
       i( S[\phi_1,\phi_{1}^{\dag} ] - S[\phi_2,\phi_{2}^{\dag} ]
     + J_1 \cdot \phi_1 - J_2 \cdot \phi_2 )
          \right]
\nonumber  \\
   &\times&
    \rho_{S}[\phi_{1i},\phi_{1i}^{\dag},\phi_{2i},\phi_{2i}^{\dag}] =
  \exp \left[ -i J^{\dag} G J \right],
\end{eqnarray}
where now the Green function $G$ acquired the $2 \times 2$ matrix
structure as
\begin{eqnarray} \label{CTPmatrix}
 G =
        \left( \begin{array}{cc}
        G_{11}         &   G_{12} \\
        G_{21}         &   G_{22}
         \end{array}      \right)
\end{eqnarray}
and
$ J^{\dag} = (J^{\dag}_1, -J^{\dag}_2) $.
For an initial thermally equilibrium state with the temperature $T=1 / \beta$,
each component of the matrix Green function in the momentum representation is given by
\begin{eqnarray}
\label{CTPGreen}
  G_{11}(p) &=&  - G^{*}_{22}(p)
  =  \theta(p_0) G_{F}(p) +
     \theta(-p_0) G^{*}_{F}(p) -2 \pi i ~\mbox{sign}(p_0) n_{B}(p) \delta(p^2-m_{S}^2),
\nonumber  \\
  G_{12}(p) &=& -2 \pi i ~\mbox{sign}(p_0) n_{B}(p) \delta(p^2-m_{S}^2),
\nonumber  \\
  G_{21}(p) &=& -2 \pi i ~\mbox{sign}(p_0) e^{\beta (p_0-\mu)} n_{B}(p) \delta(p^2-m_{S}^2),
%
\end{eqnarray}
where $G_{F}(p)= (p^2-m_{S}^2+i \epsilon)^{-1} $ is the vacuum
Feynman propagator, $n_{B}(p) \equiv (e^{\beta (p_0-\mu)} -
1)^{-1}$ is the Bose distribution function, and $\mu$ is the
chemical potential. $G$ can be diagonalized by multiplying
matrices $u_{B}$ from each side as $G = u_{B} G_{d} u_{B}^{-1}
\eta$, where
\begin{eqnarray}
 G_{d} =
        \left( \begin{array}{cc}
        G_{R}         &   0 \\
        0         &   G_{A}
         \end{array}      \right),
\end{eqnarray}
where $G_R$ and $G_A$ are retarded and advanced Green functions,
respectively. Here $u_{B}(p)$ is the thermal Bogoliubov matrix and
$\eta$ is the $2 \times 2$ Lorentz matrix which have the following
forms\cite{AurBec92,HenUme92},
\begin{eqnarray}
\label{NLSBuF}
 u_{B}(p)
   =   \sqrt{ n_{B}(p)  } e^{\beta (p_0-\mu)/2}
        \left( \begin{array}{cc}
        1    &    e^{-\beta (p_0-\mu)} \\
        1    &   1
               \end{array}      \right)
           ~\mbox{and} ~\eta
   =
        \left( \begin{array}{cc}
        1    &     0 \\
        0    &    -1
               \end{array}      \right).
\end{eqnarray}
%
Defining the thermal doublet
$ \phi = (\phi_1, \phi_2) $ and its conjugate
 $ \bar{\phi} \equiv (\phi^{\dag}_1,\phi^{\dag}_2 )\eta$,
we can write
\begin{eqnarray}
     \label{phiGphi}
 \phi^{\dag} G^{-1} \phi
 &=& \phi^{\dag} \eta u_{B} G_{d}^{-1} u_{B}^{-1} \phi
 \nonumber  \\ &=&  \bar{\phi} u_{B} G_{d}^{-1} u_{B}^{-1} \phi.
\end{eqnarray}
By redefining the field variables by global Bogoliubov transformations
$u_{B}^{-1} \phi \rightarrow \phi$ and $ \bar{\phi} u_{B}
\rightarrow \bar{\phi}$, 
one can write the partition function in Eq. 
(30) as
\begin{eqnarray}
     \label{CTP2}
 Z[J,\bar{J}]
 &=&   \int D\bar{\phi} D\phi
     \exp \left[ i  S_{0}[\phi,\bar{\phi}]
         + i \bar{J} \phi + i \bar{\phi} J \right]
    \rho_{S}[\phi_{i}, \bar{\phi_{i}}],
 \nonumber  \\
%
\end{eqnarray}
where
\begin{eqnarray}
\label{NLSBActionBar2}
 S_{0}[\phi,\bar{\phi}] &=&
 \bar{\phi} G_{d}^{-1} \phi
 \nonumber  \\ &=&
 \int d^{3}x d \omega
    \left[
 \bar{\phi} (x, \omega)
        \left(
       \omega^2  +  \vec{\partial}^2 -m_{S}^2 + i \epsilon(\omega) \eta
            \right)
 \phi(x, \omega)
     \right].
\end{eqnarray}
%
%
%
%
%
%
Here $\epsilon(\omega)\equiv \epsilon \times \mbox{sign}(\omega)$
for some infinitesimal constant $\epsilon$.
When the stochastic field $v_{S}$ has no time dependence
as in Eq. (\ref{NRHamiltonian}), the close time path action is given by
%
%
\begin{eqnarray}
\label{NLSBSv}
 S_{v_{S}}[\phi,\bar{\phi}]  &=&
 \int d^{3}x d \omega
    \left[
 \bar{\phi} (\vec{x}, \omega)
        \left(
       \omega^2  +  \vec{\partial}^2 -m_{S}^2- v_{S}(\vec{x}) + i \epsilon(\omega) \eta
            \right)
 \phi(\vec{x}, \omega)
     \right].
     \nonumber
\end{eqnarray}
%

%
%
Next we express the Green function and the partition function for
the scalar field in terms of the nonlinear sigma model.
After averaging out the partition function
in Eq. 
(28) without the source term
with respect to the stochastic potential $v_{S}$
 which obeys the following Gaussian probability
 distribution
\begin{eqnarray}
\label{NLSBPV}
 P[v_{S}] = {\cal N} \exp \left[ -\frac{1}{2}
 \int d^4 x d^4 y ~v_{S}(x) \Delta_{S}^{-1}(x-y) ~v_{S}(y) \right]
\end{eqnarray}
with the normalization constant ${\cal N}$,
we obtain the reduced action
\begin{eqnarray}
\label{NLSBreducedEA}
 Z = \langle Z[v_{S}] \rangle_{v_{S}}
 &=& \int Dv_{S} P[v_{S}] D\bar{\phi} D\phi
  \exp \left[ i S_{v_{S}}[\bar{\phi},\phi] \right] \nonumber  \\
 &=& \int D\bar{\phi} D\phi
  \exp \left[ i S_0[\bar{\phi},\phi] + i S_I[\bar{\phi},\phi] \right],
\end{eqnarray}
where
\begin{eqnarray}
\label{NLSBEA}
 S_I[\bar{\phi},\phi] &=&
   \frac{i}{2} \int d^{4}x d^{4}y
 \bar{\phi}(x) \phi(x)
 \Delta_{S}(x-y)
 \bar{\phi}(y) \phi(y).
\end{eqnarray}
Note that for a local correlation $\Delta_{S}(x-y) \sim \delta^4(x-y)$,
we obtain the effective $\phi^4$ theory similar to the one obtained from spacetime foam \cite{Foam}.

Now we extract the slow modes by the Hubbard-Stratonovich
transformation. Introducing the auxiliary bilocal matrix field
$\sigma(x,y)$ as \cite{Wegner79}
\begin{eqnarray}
\label{NLSBHS}
       e^{i S_I[\bar{\phi},\phi]  }
   &=&  \\
    \int D \sigma&
\exp&[-\frac{1}{2} \int d^4 x d^4 y \mbox{Tr}[ \sigma(x,y)
\Delta_{S}^{-1}(x-y) \sigma(y,x)  ] ] \exp \left[i S_{HS}[\sigma,\bar{\phi},\phi]  \right],
\nonumber
\end{eqnarray}
where
\begin{eqnarray}
\label{NLSBpart}
S_{HS}[\sigma,\bar{\phi},\phi] = - \int d^4 x d^4 y
~\bar{\phi}(x) \sigma(x,y) \phi(y)
\end{eqnarray}
and the trace is taken over thermal indices.
The partition function can be written as
\begin{eqnarray}
\label{NLSBZpart}
  Z
   =  \int D \sigma D\bar{\phi} D\phi
   \exp \left[ -\frac{1}{2} \int \mbox{Tr} [\sigma \Delta_{S}^{-1} \sigma]   \right]
   \exp \left[ i S_{0}[\bar{\phi},\phi] +i S_{HS}[\sigma,\bar{\phi},\phi]  \right],
\end{eqnarray}
where
$\sigma$ is Hermitian by construction, i.e. $\sigma^{\dag}(x-y) = \sigma(y-x) $.
In energy representation, Eq. 
(42) becomes
\begin{eqnarray}
\label{NLSBsma2}
S_{HS}[\sigma,\bar{\phi},\phi] =   - \int d^3 x d^3 x' \frac{d \omega}{2 \pi}\frac{d \omega'}{2 \pi}
~\bar{\phi}(\vec{x}, \omega)
     \sigma_{\omega \omega'}(\vec{x},\vec{x'})
  \phi(\vec{x'}, \omega')
\end{eqnarray}
%
and
\begin{eqnarray}
\label{NLSBActionHS2}
 S_{0}[\bar{\phi},\phi] + S_{HS}[\sigma,\bar{\phi},\phi]
 &=& \\
\int d^3 x d^3 x' \frac{d \omega}{2 \pi}\frac{d \omega'}{2 \pi} ~\bar{\phi}(\vec{x}, \omega)
 \left[
  \left(
     \omega^2
   \right.
   \right.
     &+&
   \left.
   \left.
     \vec{\partial}^2 -m_{S}^2+ i \epsilon(\omega) \eta
  \right)
 \delta(\vec{x} - \vec{x'}) \delta(\omega - \omega')
 - \sigma_{\omega \omega'}(\vec{x},\vec{x'})
 \right]
\phi(\vec{x'}, \omega').
 \nonumber
\end{eqnarray}

After integrating out $\bar{\phi}$ and $\phi$, we obtain
\begin{eqnarray} \label{NLSBZsigma}
  Z
   &=&  \int D\sigma
     \exp \left[ -\frac{1}{2} \int \mbox{Tr} [\sigma \Delta_{S}^{-1} \sigma] \right]
\\
  \times &\exp& \left[ -
  \int d^3 x d^3 x' \frac{d \omega}{2 \pi}\frac{d \omega'}{2 \pi}
 \mbox{Tr} \log
 \left[
  \left(
    \omega^2  +  \vec{\partial}^2 -m_{S}^2+ i \epsilon(\omega) \eta
  \right)
 \delta(\vec{x} - \vec{x'}) \delta(\omega - \omega')
 - \sigma_{\omega \omega'}(\vec{x},\vec{x'})
 \right]
               \right].
     \nonumber
\end{eqnarray}
The nonlinear sigma model is commonly used in many different areas
in physics to study collective excitations and the dynamical
symmetry breaking property of the system.
Nevertheless it has not been studied previously in
the closed time path method in detail. 
When applied to the disordered
systems, it is required that all the fermion loop diagrams will
cancell in order to include the effects of elastic scattering due
to impurity which carries no energy. Such techniques as replica
formalism \cite{Wegner79} or supersymmetric extension
\cite{Efetov97} are commonly used for this purpose. A general
class of real time path ordered methods
\cite{SimUme83,NieSim84,LanWee87,HorSch90,CLN99} is also known to
have such a property owing to the energy integral.
The closed time path formalism employed here has the
advantage compared to other methods in that it is naturally
extensible to the nonequilibrium setting.
The trace part in Eq. (\ref{NLSBZsigma}) can be expanded in terms of 
the $\sigma$ field as
\begin{eqnarray}  \label{NLSBtracelog}
  \mbox{Tr}& \log &\{
   \left(
   \omega^2  +  \vec{\partial}^2 -m_{S}^2
    + i \epsilon(\omega) \eta
   \right) \delta(\omega - \omega')
    \delta(\vec{x} - \vec{x'})
 - \sigma_{\omega \omega'}(\vec{x},\vec{x'})
   \}
   \nonumber  \\
&=&
 \mbox{Tr} \log \left[
 G^{-1} - \sigma   \right]
 \nonumber  \\
 &=&
 \mbox{Tr} \log \left[
 G^{-1}  \right]
 +
\sum_{n=1} \frac{(-1)^n}{n} \mbox{Tr} \left[
G \sigma
 \right]^n,
\end{eqnarray}
where $ G(x, x')
\equiv ( \Box + m_{S}^2)^{-1}$ is the free boson propagator.

Now we assume that the spatial correlation of disorder decays
sufficiently fast so that the $\sigma$ field can be treated as a
local field variable in space. 
This implies that the stochastic potential $v_{S}$ has 
the time-independent form
\begin{eqnarray} \label{vdistribution}
P[v_{S}] = {\cal N} \exp \left[ -\frac{1}{2u_{S}} \int d^3 x
~v_{S}^2(\vec{x}) \right].
\label{sigma}
\end{eqnarray}
In this case the partition function is reduced to
\begin{eqnarray} \label{localNLSmodel}
  Z
   &=&  \int D\sigma
   \exp \left[ -\frac{1}{2u_{S}} \int d^3 x \mbox{Tr} \sigma^2(\vec{x}) 
\right]
\\
  &\times& \exp \left[ - \int d^3 x  \frac{d \omega}{2 \pi}\frac{d \omega'}{2 \pi}  \mbox{Tr} \log
   \{
     \left(
  \omega^2  +  \vec{\partial}^2 -m_{S}^2
+ i \epsilon(\omega) \eta
   \right) \delta(\omega - \omega')
 - \sigma_{\omega \omega'}(\vec{x})
    \}  \right]. \nonumber
\end{eqnarray}
The equation of motion obtained from Eq. (\ref{localNLSmodel}) corresponds to
the one obtained from the coherent potential approximation:
\begin{eqnarray}  \label{CPA}
 \sigma_{\omega \omega'}(\vec{P})
=
\frac{u_{S}}{
   \omega^2  - \vec{P}^2 -m_{S}^2
    + i \epsilon(\omega) \eta
    - \sigma_{\omega \omega'}(\vec{P})
   }.
\end{eqnarray}
The real part of the $\sigma$ field gives the mass and frequency
renormalization and will not be considered hereafter. The
imaginary part of the $\sigma$ field $\sigma_{I}(\omega)$ gives
the scattering rate. Writing the homogeneous solution of Eq.
(\ref{CPA}) as $\sigma_{\omega \omega'}(\vec{P}) = -i
\sigma_{I}(\omega) \mbox{sign}(\omega)
  \delta(\vec{P}) \delta(\omega - \omega')$,
one obtains the relation
\begin{eqnarray}
\sigma_{I}(\omega) = \frac{u_{S} \pi N(\omega) }{2 |\omega|},
 \label{relaxationtime}
\end{eqnarray}
  where $N(\omega)$ is the density of states of the scalar field.

The quadratic term in Eq. (\ref{NLSBtracelog})
can be written as
\begin{eqnarray}  \label{NLSBSSterm}
\frac{1}{2}  \int d^3 x_1 d^3 y_1 d^3 x_2 d^3 y_2 \frac{d \omega}{2 \pi}\frac{d \omega'}{2 \pi}
 \mbox{Tr} \left[
G^{aa}(\vec{x}_1, \vec{y}_1, \omega )  \sigma^{ab}_{\omega \omega'}(\vec{y_1},\vec{x_2})
G^{bb}(\vec{x}_2, \vec{y}_2, \omega' )  \sigma^{ba}_{\omega' \omega}(\vec{y_2},\vec{x_1})
 \right].
\end{eqnarray}
In the momentum representation, the above expression becomes
\begin{eqnarray}  \label{NLSBssmomentum}
\frac{1}{2}
   \int \frac{d^4 P}{(2 \pi)^4} \frac{d^4 k}{(2 \pi)^4}
   \mbox{Tr}
   \left[
       G^{bb}(k + \frac{P}{2})   \sigma^{ba}_{k}(-P)
       G^{aa}(k - \frac{P}{2})   \sigma^{ab}_{k}(P)
   \right].
\end{eqnarray}
The Fourier component of the $\sigma$ field is defined as
\begin{eqnarray}  \label{sigmamomentum}
   \sigma(x,y) = \int \frac{d^4 P}{(2 \pi)^4} \frac{d^4 k}{(2 \pi)^4}
   \sigma_{k}(P) e^{-i k r} e^{-i P X},
\end{eqnarray}
where $r \equiv x-y$ and $X \equiv (x+y)/2$.
Thus we obtain the kinetic term of the $\sigma$ field in Eq. (\ref{NLSBZsigma})
\begin{eqnarray}  \label{NLSBssmomentum2}
\frac{1}{2}
   \int \frac{d^4 P}{(2 \pi)^4} \frac{d^4 k}{(2 \pi)^4}
   \mbox{Tr}
   \{
     \sigma^{ba}_{k}(-P)
    \left[ \frac{1}{u_{S}} -
        G^{aa}(k - \frac{P}{2})
        G^{bb}(k + \frac{P}{2})
    \right]
     \sigma^{ab}_{k}(P)
   \}.
\end{eqnarray}
If the spatial dependence in the $\sigma$ field is local, the
$\sigma$ field is independent of the momentum $\vec{k}$, i.e.  $
\sigma_{k}(P) = \sigma_{k_0}(P)$ and we can integrate out
$\vec{k}$ and obtain
\begin{eqnarray}  \label{NLSBk0}
\frac{1}{2}
   \int \frac{d^4 P}{(2 \pi)^4} \frac{d k_0}{2 \pi}
   \mbox{Tr}
   \{
     \sigma^{ba}_{k_0}(-P)
    \left[ \frac{1}{u_{S}} -
     \int_{\vec{k}}  G^{aa}(k - \frac{P}{2})
       G^{bb}(k + \frac{P}{2})
    \right]
    \sigma^{ab}_{k_0}(P)
   \}.
\end{eqnarray}
By expanding inside the bracket in Eq. (\ref{NLSBk0})
with respect to $P_0$ and $\vec{P}$,
one can show that, in the limit $\sigma_{I}(k_0) >> k_0 P_0$,
the off diagonal term of the free propagator of the $\sigma$ field
in the thermal indices gives the massless excitation
which has the form of the diffusion propagator \cite{McKSto81}
%
\begin{eqnarray} \label{NLSBpropq}
\langle
 \sigma^{ab}_{k_0}(P)
 \sigma^{ba}_{k_0}(-P)
\rangle = \frac{2 \sigma_{I}^{2}(k_0)}{\pi N(k_0)  }
          \frac{1}{D(k_0) \vec{P}^2 - i P_0 \eta^{aa}},
\end{eqnarray}
where $D(k_0)$ is the diffusion constant. Diagramatically this
form is obtained by including all the ladder diagrams in the
particle-hole propagator.
 The diffusion constant is related to the
dc conductivity $C_0$ through the Einstein relation:
 $C_0 = N(k_0) D(k_0) $.
The diagonal term in Eq. (\ref{NLSBk0}) gives the massive
excitation that can be integrated out. However, it does not
contribute directly to the infrared divergence responsible for the
universal behaviour of the conductance fluctuations and will not
be considered in this work.

In this $U(2)$ nonlinear sigma model, the matrix field $\sigma$
takes its value in the coset space $U(2) / U(1) \times U(1)$.
After making a transformation
 $\sigma \rightarrow V^{-1} \sigma V$ with $V = \theta(\omega)1 + \theta(-\omega)
\sigma_2$, where matrices $1$ and $\sigma_2$ act on the thermal
indices, the saddle point solution changes its form as
$\mbox{sign}(\omega) \eta \rightarrow \eta$. In this
representation, the field $\sigma$ around the saddle point can be
parametrized as \cite{Hikami81}
\begin{eqnarray} \label{stoV}
 \sigma &=&
     \left(    \begin{array}{cc}
       \sqrt{1 - q q^{\dag}} & q  \\
       q^{\dag}              & - \sqrt{1 - q^{\dag} q}
                         \end{array}    \right)
\nonumber  \\ &=&
     \left(    \begin{array}{cc}
        1 & 0            \\
        0 & -1           \end{array}    \right)
 +   \left(    \begin{array}{cc}
        0 & q            \\
        q^{\dag} & 0     \end{array}    \right)
 -   \frac{1}{2}
     \left( \begin{array}{cc}
        q q^{\dag} & 0   \\
        0 & -q^{\dag} q  \end{array}    \right)
       \cdots.
\end{eqnarray}
The matrix fields $q=q_{\omega\omega'}$ and
$q^{\dag}=q^{\dag}_{\omega'\omega}$ carry two frequency indices.
Inserting this expression in Eq. (\ref{NLSBk0}),
we obtain the form of the free propagator of $q$ field as
\begin{eqnarray} \label{propq}
\langle
 q_{k_0}(P)
 q^{\dag}_{k_0}(-P)
\rangle &=& \frac{2}{\pi N(k_0) }
          \frac{1}{D(k_0) \vec{P}^2 - i P_0 }.
\nonumber  \\
%
\end{eqnarray}
 $\sigma_{I}(\omega)$ in the numerator was absorbed in the redefinition of $q$.
Thus the partition function for the linear part of the sigma model
with respect to $q$ field is
\begin{eqnarray} \label{NLSs}
  Z
   =  \int  D q^{\dag} Dq
     \exp \{ -\pi
   \int \frac{d^4 P}{(2 \pi)^4} \frac{d k_0}{2 \pi}
 N(k_0)
q^{\dag}_{k_0}(-P)
 \left[
    D(k_0) \vec{P}^2 - i P_0
 \right]
    q_{k_0}(P)
         \}.
\end{eqnarray}
Higher order vertex terms will follow corresponding to the
expansion in Eq. (\ref{stoV}).
 The fields $q$ and $q^{\dag}$ are free from the constraint and
take all possible values. Time dependence in the stochasticity can
be ascribed to the intrinsic property of the effective medium and
be treated as the frequency dependent random potential.
 The modification associated with this change can be
absorbed in the diffusion constant as the term proportional to
$\partial v_{S}/ \partial \omega$. If this correction is too
large, a different approach is required. The usual prescription to
account for the effect of inelastic scattering is to include
inelastic scattering rate $\Delta$ in the denominator of the
diffusion propagator so that we replace $ D(k_0) \vec{P}^2 - i P_0$
in Eq. (\ref{NLSs}) simply by $ D(k_0) \vec{P}^2 -i P_0 + \Delta$.
$\Delta$ will set the time scale beyond
which the phase memory of a scattered wave is lost and the transport behavior
becomes classical.

\subsection{Mesoscopic fluctuations of scalar fields}

%
\begin{figure}[h]
  \begin{center}
\epsfxsize=.4\textwidth \epsfbox{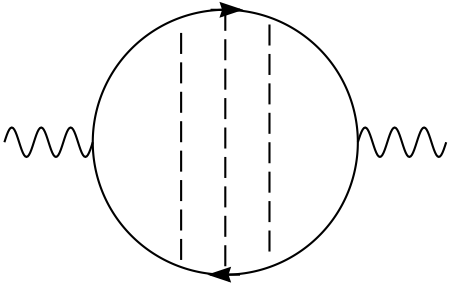}
  \end{center}
\caption{The Feynman diagram for the conductance. The metric
fluctuations are included as a ladder contribution which is
represented by the dotted line.} \label{fig2}
\end{figure}
The conductivity associated with the Noether current in the
presence of the external field with a frequency $\kappa$ can be
written by the current-current correlation function by the Kubo
formula \cite{Kubo57} as
\begin{equation}\label{NLSBnoncond}
C_{\kappa}(\vec{x},\vec{y}) \equiv \frac{\pi}{\kappa}
 \int_{-\infty}^{\infty} d \omega \Omega_{\kappa}(\omega)
 \sum_{mn} j_{mn}(\vec{x})
j_{nm}(\vec{y}) \delta(\omega + \kappa - \omega_n) \delta(\omega -
\omega_m),
\end{equation}
where $j_{mn} \equiv i \phi^{\dag}_m
\stackrel{\leftrightarrow}{\partial} \phi_n$ is the Noether
current expressed by two energy eigenstates and
$\Omega_{\kappa}(\omega)$ is a smearing function that depends on
the characteristics of the system and the environment. Near
equilibrium, it can be written as $\Omega_{\kappa}(\omega) =
\rho_{S}[\omega]-\rho_{S}[\omega+\kappa]$, where
$\rho_{S}[\omega]$ is the initial density matrix. If the metric
fluctuations are independent of temperature as we assume, the
effect of temperature on the conductivity only arises from this
term.
 Here we assume
that $\Omega_{\kappa}(\omega)$ is normalized such that $\int d
\omega \Omega_{\kappa}(\omega) = \kappa$. In Figure 2, the
conductance is represented by the Feynman diagram. In the leading
order weak disorder expansion, averaging over disorder is
taken into account by including all the ladder diagrams. The
conductivity can be written in the more familiar form in terms of
Green functions as \cite{FisherLee81}
\begin{eqnarray}\label{NLSBnoncondG}
C_{\kappa}(\vec{x},\vec{y}) & & \equiv \\
 - \frac{1}{4 \pi \kappa}
 \int_{-\infty}^{\infty} d \omega & & \Omega_{\kappa}(\omega)
 \left[ G_R(\vec{x},\vec{y},\omega) -  G_A(\vec{x},\vec{y},\omega) \right]
\stackrel{\leftrightarrow}{\partial}_x
\stackrel{\leftrightarrow}{\partial}_y
 \left[ G_R(\vec{y},\vec{x},\omega+\kappa)
     -  G_A(\vec{y},\vec{x},\omega+\kappa) \right].
     \nonumber
\end{eqnarray}
%
With the expression of Green functions in terms of thermal fields,
\begin{eqnarray} \label{NLSBG+-}
G_{R}(x,y) &\equiv&
- i \theta(x_0 - y_0) \langle
                      \left[ \hat{\phi}(x), \hat{\phi}^{\dag}(y)
                      \right]
              \rangle
= i \langle \phi_1(x) \phi^{\dag}_1(y) \rangle, \nonumber  \\
 G_{A}(x,y)
&\equiv& i \theta(y_0 - x_0) \langle
                     \left[ \hat{\phi}(x), \hat{\phi}^{\dag}(y)
                     \right]
                    \rangle
= - i \langle \phi_2(x) \phi^{\dag}_2(y) \rangle,
\end{eqnarray}
we write Eq. (\ref{NLSBnoncondG}) as
\begin{eqnarray}\label{NLSBnoncondPhi}
C_{\kappa}(\vec{x},\vec{y})& &\equiv \nonumber  \\
  \frac{-1}{4 \pi
\kappa}  \int_{-\infty}^{\infty} d \omega & &
\Omega_{\kappa}(\omega)
\sum_{abcd}  \langle \phi^{a}(\vec{x},\omega)
\stackrel{\leftrightarrow}{\partial}_x \phi^{\dag
b}(\vec{x},\omega + \kappa) \phi^{c}(\vec{y},\omega + \kappa)
\stackrel{\leftrightarrow}{\partial}_y \phi^{\dag
d}(\vec{y},\omega) \rangle.
\end{eqnarray}
Note that even though the thermal indices in the sum run over all
possible values, only pairwise equal terms contribute to the
conductivity.

 This expression
can be obtained directly from the partition function in terms of
the functional derivative by introducing the external source term
$\vec{A}^{\kappa}$ in the form
\begin{equation}\label{NLSBsource}
S[A,\bar{\phi},\phi] = i \int d^3 x
\frac{d \omega}{2 \pi}\frac{d \omega'}{2 \pi}
~\bar{\phi}(\vec{x}, \omega) \vec{A}^{\kappa}_{\omega}(\vec{x})
 \cdot \stackrel{\leftrightarrow}{\partial} \delta_{\kappa}
  \phi(\vec{x}, \omega'),
\end{equation}
with
\begin{eqnarray}\label{Lmatrix}
\delta_{\kappa} \equiv
\delta(\omega' - \omega + \kappa)
         \left( \begin{array}{cc}
    1 & 1 \\
    -1 & -1
               \end{array}      \right).
\end{eqnarray}
The matrix in $\delta_{\kappa}$ acts on the thermal indices.
Note that $\delta_{\kappa}$ is nilpotent, i.e. $\delta_{\kappa}^2
=0$. The nonlocal conductivity is given by
\begin{equation}\label{NLSBcondderiv}
C_{\kappa}(\vec{x},\vec{y}) = \frac{- \pi}{\kappa V}
 \frac{\delta^2 W[A]}
 {\delta \vec{A}^{\kappa}(\vec{x}) \delta \vec{A}^{-\kappa} (\vec{y}) },
%
%
%
%
\end{equation}
where $V$ is the spatial volume of the system and the integral
over energy indices are understood.
In the presence of the source term, the partition function for
the matrix field $\sigma$ in Eq. (\ref{NLSBZsigma}) becomes
\begin{eqnarray} \label{NLSBZsigma2}
Z[A] &=& e^{i W[A]} = \int D\sigma
  \exp \left[ -\frac{1}{2} \int \mbox{Tr}
       \left[\sigma \Delta_{S}^{-1} \sigma \right]
       \right]
\exp
 \left[
  \int d^3 x d^3 x' \frac{d \omega}{2 \pi} \frac{d \omega'}{2 \pi}
 \right.
\\
  &\mbox{Tr}&
 \left.
\log \{
    \left[
       \left(
              \omega^2  +  \vec{\partial}^2 -m_{S}^2
              + i \epsilon(\omega) \eta
       \right) \delta(\omega - \omega')
        + i \vec{A}^{\kappa}_{\omega} \cdot \stackrel{\leftrightarrow}{\partial} \delta_{\kappa}
    \right] \delta(\vec{x} - \vec{x'})
       - \sigma_{\omega \omega'}(\vec{x},\vec{x'})
     \}
 \right]. \nonumber
\end{eqnarray}
%

Making use of the gauge symmetry in Eq. (\ref{NLSBZsigma2}), for a
constant vector field $\vec{A}^{\kappa}$, the source term in above 
can be generated by the following gauge
transformation
\begin{eqnarray}
      \phi(\vec{x})  &\rightarrow&
      e^{i \vec{x} \cdot \vec{A}^{\kappa} \delta_{\kappa}}  \phi(\vec{x}), \nonumber  \\
 \bar{\phi}(\vec{x}) &\rightarrow& \bar{\phi}(\vec{x})
      e^{-i \vec{x} \cdot \vec{A}^{\kappa} \delta_{\kappa}}.
      \label{NLSBgaugecoupl}
\end{eqnarray}
Correspondingly, the Green function $G(\vec{x}, \vec{y}, \omega)$
and the $\sigma$ field transform as
\begin{eqnarray}
  G(\vec{x}, \vec{y})
 \rightarrow  U^{-1}(\vec{x})
    G(\vec{x}, \vec{y})
    U(\vec{y})
      \label{NLSBosonPropagator}
\end{eqnarray}
and
\begin{eqnarray}
\sigma(\vec{x}, \vec{y}) \rightarrow
U^{-1}(\vec{x}) \sigma(\vec{x}, \vec{y}) U(\vec{y}),
\label{sigmatransform}
\end{eqnarray}
where $U(\vec{x}) \equiv e^{-i \vec{x} \cdot \vec{A}^{\kappa}
\delta_{\kappa}} $.
This gauge symmetry will induce the gauge coupling in the
effective Lagrangian through the covariant derivative
$\vec{\nabla} \equiv \vec{\partial} + i \vec{A}^{\kappa}
\delta_{\kappa}$:
\begin{eqnarray}
 &\mbox{Tr}& \left( \vec{\nabla} \sigma \vec{\nabla} \sigma \right)
  \nonumber  \\
 &=&
  \mbox{Tr} \left( \vec{\partial} \sigma \vec{\partial} \sigma \right)
 + 2 i  \mbox{Tr} \left( \vec{A}^{\kappa} \delta_{\kappa} \sigma \vec{\partial} \sigma \right)
 - \mbox{Tr} \left( \vec{A}^{\kappa_{1}} \delta_{\kappa_{1}} \sigma
 \vec{A}^{\kappa_{2}} \delta_{\kappa_{2}} \sigma \right).
\label{NLSBAexpansion}
\end{eqnarray}

Now we obtain the expression of the conductivity in terms of the $\sigma$ field:
\begin{eqnarray}\label{NLSBcond}
C_{\kappa} &=& \frac{-\pi}{ \kappa V}
 \frac{\delta^2 W[A]}{\delta \vec{A}^{\kappa} \delta \vec{A}^{-\kappa} } \\
%
%
%
%
&=& \frac{-\pi}{ \kappa V} \left[
 \int d^3 x_1 \langle \mbox{Tr}
\left[ \delta_{\kappa} \sigma(x_1) \delta_{-\kappa} \sigma(x_1)
\right] \rangle
 -
 \int d^3 x_1 d^3 x_2
  \langle
\mbox{Tr} \left[\delta_{\kappa} \sigma(x_1) \vec{\partial}
\sigma(x_1) \right] \mbox{Tr} \left[ \delta_{-\kappa} \sigma(x_2)
\vec{\partial} \sigma(x_2) \right] \rangle \right]    \nonumber
\end{eqnarray}
and its fluctuation:

\begin{eqnarray}\label{NLSBcondfl}
{C}^{2}_{\kappa_{1} \kappa_{2}} &=& \frac{\pi^2}{V^2  \kappa_{1}
\kappa_{2} }
 \frac{\delta^4 W[A]}
 {\delta \vec{A}^{\kappa_{1}} \delta \vec{A}^{-\kappa_{1}}
  \delta \vec{A}^{\kappa_{2}} \delta \vec{A}^{-\kappa_{2}}  }
\nonumber  \\
&=& \frac{1}{V} \left[
           C^{(1)} - C^{(2)} + C^{(3)} \right],
\end{eqnarray}
where
\begin{eqnarray}\label{C1}
 C^{(1)} &=&
 \frac{2 \pi^2}{\kappa_{1} \kappa_{2} V}
 \int d^3 x_1 d^3 x_2 \langle \mbox{Tr} \left[\delta_{\kappa_{1}}
\sigma(\vec{x}_1) \delta_{-\kappa_{1}} \sigma(\vec{x}_1)\right] \mbox{Tr}
\left[ \delta_{\kappa_{2}} \sigma(\vec{x}_2)
\delta_{-\kappa_{2}} \sigma(\vec{x}_2) \right]
\rangle,
\nonumber  \\
%
%
%
 C^{(2)} &=&
  \frac{4 \pi^2}{\kappa_{1} \kappa_{2} V}
  \int d^3 x_1 d^3 x_2 d^3 x_3
 \langle
\mbox{Tr} \left[\delta_{\kappa_{1}} \sigma(\vec{x}_1) \vec{\partial} \sigma(\vec{x}_1) \right]
\mbox{Tr} \left[\delta_{-\kappa_{1}} \sigma(\vec{x}_2) \vec{\partial}
\sigma(\vec{x}_2)\right]
\mbox{Tr} \left[\delta_{\kappa_{2}} \sigma(\vec{x}_3)
\delta_{-\kappa_{2}} \sigma(\vec{x}_3) \right] \rangle,
\nonumber  \\
%
%
\mbox{and} \nonumber  \\
%
%
 C^{(3)}
 &=&  \frac{\pi^2}{\kappa_{1} \kappa_{2} V}
 \int d^3 x_1 d^3 x_2 d^3 x_3 d^3 x_4
\nonumber  \\
 \langle & \mbox{Tr} &
\left[\delta_{\kappa_{1}} \sigma(\vec{x}_1) \vec{\partial} \sigma(\vec{x}_1) \right]
\mbox{Tr}
\left[\delta_{-\kappa_{1}} \sigma(\vec{x}_2) \vec{\partial} \sigma(\vec{x}_2) \right]
\mbox{Tr} \left[\delta_{\kappa_{2}} \sigma(\vec{x}_3) \vec{\partial}
\sigma(\vec{x}_3) \right]
\mbox{Tr} \left[\delta_{-\kappa_{2}} \sigma(\vec{x}_4)
\vec{\partial} \sigma(\vec{x}_4) \right] \rangle. \nonumber
\end{eqnarray}
Here we further included the energy indices in the definition of
the trace as 
$\mbox{Tr}\left( {\cal O} \right) \equiv \pi \int d
k_0 / (2 \pi)^2 N(k_0) D(k_0) \int d P_0 \sum_{a} {\cal
O}^{aa}(k_0,P_0)$. We are interested in the dc conductivity
evaluated in the limit $\kappa_{1}$,$\kappa_{2} \rightarrow 0$.
Note that $\Omega_{\kappa}(\omega)$ is generally a peak function
which describes a wave packet peaked around the specific mode
$\omega=\omega_0$. Furthermore,
if we assume for simplicity that it is given by
the step function: $\Omega_{\kappa}(\omega) = 1$ (for $\omega_0 <
\omega < \omega_0 + \kappa$) and 0  (otherwise),
then taking the dc limit $\kappa \rightarrow 0$ extracts the
particular mode $\omega_{0}$. The effect of finite temperature $T$
can be viewed as an additional smearing due to the width of
$\Omega_{\kappa}(\omega)$. From Appendix B, in the dc limit, the
formula above gives the simple result
\begin{eqnarray}\label{Csum}
C_{0}^{2} &=& \lim_{\kappa_{1},\kappa_{2} \rightarrow
0}{C}^{2}_{\kappa_{1} \kappa_{2}} = \frac{c}{V M_{IR}},
\end{eqnarray}
where $c=7.295 \cdots$ is a constant and $M_{IR}$ is the infrared
cutoff of the momentum integral. In Figure 3, the corresponding
Feynman diagrams for the conductance fluctuations are given. Other
diagrams that contain crossed diagrams contribute as higher order
terms in the weak disorder expansion. Here we assume that the
region with the fluctuating metric is restricted in the finite
cube with the edge length $L$ and that the rest of the spacetime
is flat. This enables us to handle the problem as a scattering
process. Taking $L$ as the parameter that lies in the {\it
mesoscopic scale}, the analogy with the electric circuit becomes
elucidated. If we use the relation $C_0 = g L^{2-d}$ for the
conductivity in $d$ dimension for the dimensionless conductance
$g$, we obtain the conductance fluctuations in terms of the
conductivity fluctuations as $g^2 = C^2 L^{2d-4}$. Then from Eq.
(\ref{Csum}), in three dimension, using $M_{IR} \sim \pi/L$ and we
obtain
\begin{eqnarray}
  g^2 \sim 2.322 \cdots.
      \label{ConductanceFluc}
\end{eqnarray}
This value is universal in the sense that it is independent of the
amount of stochasticity initially assumed and the size of the
stochastic region \cite{UCF,AKL91}. The conductance is also
directly related to the transmission matrix. Indeed one can show
that the conductance measures the intensity of wave transmission
\cite{FisherLee81} and the fluctuations of conductance correspond
to the fluctuations of wave intensity.
%
\begin{figure}[h]
  \begin{center}
\epsfxsize=.9\textwidth \epsfbox{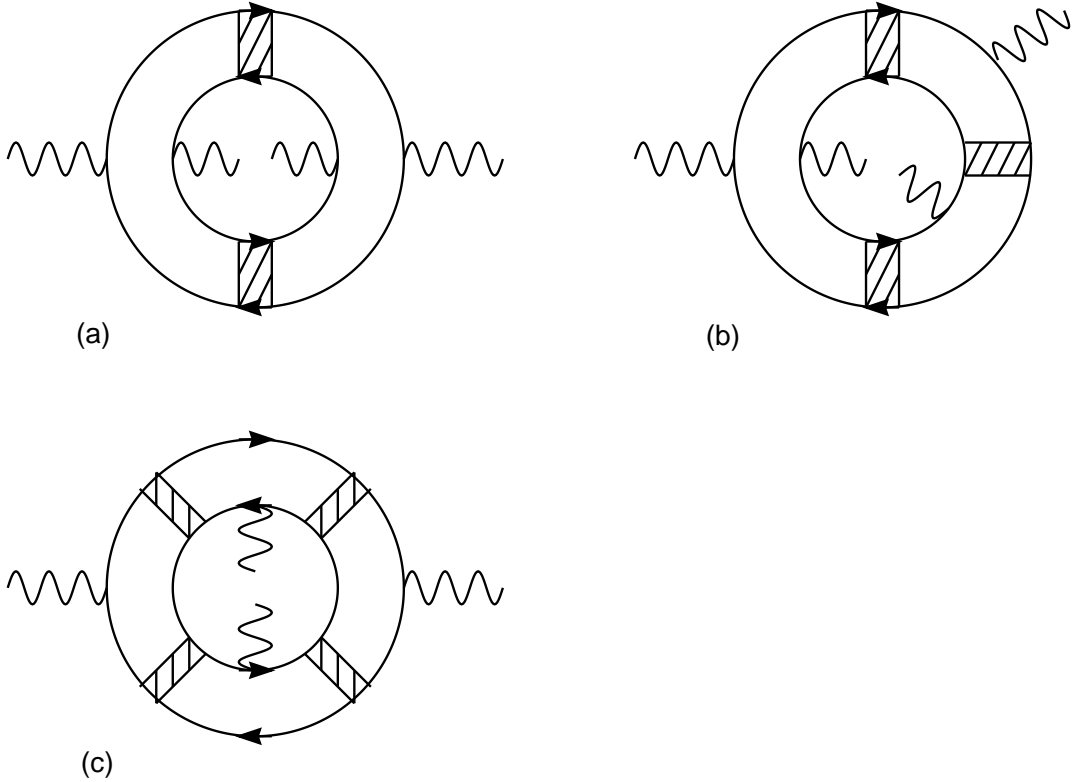}
  \end{center}
\caption{The conductance fluctuations are represented by Feynman
diagrams. The shaded regions are diffusion propagators.
(a),(b), and (c) contain two, three, and four diffusions and
correspond to $C^{(1)}$, $C^{(2)}$, and $C^{(3)}$ in Eq. (\ref{NLSBcondfl}),
respectively.
} \label{fig3}
\end{figure}
\section{Discussion}

In this paper, we showed the analogy between the field propagation
in Minkowski spacetime with a small stochasticity in the metric
and the wave in disordered systems.
While the electromagnetic
field propagation in a stochastic spacetime is similar to that in
a random media, the scalar and spinor field propagation was shown to be
similar to the electron in a disordered potential.
Both cases can be treated similarly, however, the following difference
should be noted.
In the former, the field remains massless and the randomness affects
the refraction property of light differently depending on the frequency of
 the wave. In particular, low energy scattering is suppressed;
In the latter,  a random mass causes scattering with any energy.
Mesoscopic fluctuations associated with wave propagation were characterized
by the nonlinear sigma model in the closed time path method. We
introduced the collective fields by the Hubbard-Storatonovich
transformation and integrated out the fundamental field variables
and obtain the nonlinear sigma model written in terms of the
collective fields only. The conductivity and its fluctuations were
expressed by these fields. For the time independent or slowly
dependent stochasticity, the fluctuations of the dc conductivity
were shown to be universal and of order unity. The origin of this
universality is traced back to the infrared divergence due to the
Nambu-Goldstone boson which appears as a result of symmetry
breaking.


%
Although the induced effects on the propagation of waves in the
presence of metric fluctuations are themselves of theoretical and
observational importance as long as the backreaction of the matter
fluctuations is small, a self-consistent treatment is necessary
for Planck scale processes\cite{CamHu98}. This line of consideration is
important particularly of the metric fluctuations produced in the
cosmological processes.

Relativistic quantum field theoretic calculation of the transport
coefficients has been developed during the past decade
\cite{HST84,JeoYaf96,CHR99}. The electrical conductivity in the early
universe controlls the generation of primordial magnetic field
which is believed to be the origin of the strong magnetic fields
presently observed in spiral galaxies \cite{BayHei97,BdVS99}. The
method developed in Sec. 3 also gives the field theoretic basis
for the study of, for example, mesoscopic fluctuations due to
random magnetic fields.

%
The coarse graining of microscopic degree of freedom
necessarily induces the nonlocal correlation in the stochastic
fields.
Moreover, unitarity in the whole system guarantees the relation between
the dissipation kernel
and the noise kernel in the form of the fluctuation-dissipation theorem
upon coarse graining.
In the present work, the nonlocal, noncommutative origin of
the stochastic fields and the effect of dissipation are ignored and
only the classical aspects are considered.
Possible manifestation of the quantum nature of
underlying microscopic gravitational dynamics
in the mesoscopic effects remains to be clarified \cite{Garay98}.
Fluctuating metric is also relevant to infer possible decoherence
effects in the quantum interference of propagating particles
\cite{RosSan91,DioHal98,PowPer98}.
 The closed time path method 
gives a suitable framework to discuss such effects. Our
results of conductance fluctuations assume that the time scale of
metric fluctuations is relatively long.
 In such a case, the time
scale of fluctuations appears in the coherence time scale to
restrict the validity of the arguments and diagrammatic
calculations based on the coherence between different modes beyond
this time scale. This prescription is quite successful in
explaining many mesoscopic experimental results such as electron
scattering in a helium gas. 
Thus, the heavy defects randomly created in
the phase transition in early universe can be the origin of such
fluctuations of the metric.
%


Since, we have not specified the origin of stochasticity in
the metric in this work, 
explicit derivation of such stochasticity from the
fundamental model of gravity is desired.
Spacetime uncertainty proposed in the context of string theory
\cite{LiYon97}
 may have similar effects
as discussed in this work on low energy physics. Branes or other
solitonic objects that appear in string theory acquire heavy mass
in the weak string coupling limit and become another possible source of
stochasticity. Along with the possible decoherence associated with
fluctuating metric, mesoscopic effects treated here may have an
observable consequence on future experiments
\cite{PerStr97,AEMNS98}. These microscopic origins of metric
fluctuations are intrinsically beyond the validity of
semiclassical gravity. Therefore the metric fluctuations
introduced as the modification of semiclassical Einstein equation
in this paper possibly capture the essential effects of near
Planck scale physics on the sub-Planckian scale physics
effectively while the self-consistency based only on the
conventional semiclassical Einstein equation may not have a
predictive power on such a phenomenon. Clarifying the difference
between these approaches needs more careful study.
Effects of metric fluctuations in cosmological and black hole
spacetimes are considered by many authors, for example, in
\cite{CCV97,York83,Ford97,HRS98,BFP98,Sorkin99}. To identify how
the mesoscopic effects discussed in this paper manifest themselves
 in such curved
spacetimes is of particular interest. These directions are currently in
progress.

%
%
\noindent {\bf Acknowledgement} The author thanks Prof. B. L. Hu
for various discussions which motivated the present work, Prof. V.
Frolov for useful comments, and Prof. D. Page for explaining his
relevant work on the spacetime foam.
%
\appendix
\renewcommand{\theequation}{\thesection\arabic{equation}}

\section{Mesoscopic fluctuations of Dirac fields}
\label{app:Mesoscopic fluctuations of Dirac fields}
\setcounter{equation}{0} In this appendix we consider the
quantization of the Dirac field which obeys the equation of motion
in Eq. (\ref{DiracStart2}). The action for this system has the
following form,
\begin{eqnarray}
\label{ActionBar}
 S_{v_{D}}[\psi,\bar{\psi}]=
 \int d^{4}x
 \bar{\psi}(x)
        \left(
       i \not{\!\partial} - m_{D} - v_{D}(x)
        \right)
 \psi(x).
\end{eqnarray}
This action is invariant under the global $U(1)$ gauge
transformation
\begin{eqnarray}
      \label{Diracglobalgauge}
      \psi(x)  &\rightarrow& e^{i \eta}   \psi(x), \nonumber  \\
 \bar{\psi}(x) &\rightarrow& \bar{\psi}(x) e^{-i \eta},
\end{eqnarray}
and the corresponding Noether current is
\begin{eqnarray}
      \label{f302}
         J_{\mu}(x) =
      i \bar{\psi}  \gamma_{\mu} \psi.
\end{eqnarray}

In the absence of the stochastic field, the closed time path
partition function has the form corresponding to Eq.
(30)
\begin{eqnarray}
     \label{FermionCTP}
Z[J,J^{\dag}]
   &=&
    \int d\psi_{f} d\bar{\psi_{f}}
    D\psi_1 D\bar{\psi_{1}} D \psi_2 D \bar{\psi_{2}}
     \exp \left[
       i( S[\psi_1,\bar{\psi_{1}} ] - S[\psi_2,\bar{\psi_{2}} ]
     + J_1 \cdot \psi_1 - J_2 \cdot \psi_2 )
          \right]
\nonumber  \\
   &\times&
    \rho_{D}[\psi_{1i},\bar{\psi_{1i}},\psi_{2i},\bar{\psi_{2i}}]
 =
  \exp \left[ -i J^{\dag} S J \right],
\end{eqnarray}
where $J \cdot \psi \equiv \int d^4 x \left[ J^{\dag}(x) \psi(x) +
\psi^{\dag}(x) J(x) \right]$. The matrix Green function $S$ has
the form
\begin{eqnarray}
          \label{FermionCTPmatrix}
 S =
        \left( \begin{array}{cc}
        S_{11}         &   S_{12} \\
        S_{21}         &   S_{22}
         \end{array}      \right),
\end{eqnarray}
whose components are
\begin{eqnarray}
\label{FermionCTPGreen}
    S_{11}(p) &=& - S^{*}_{22}(p)
\nonumber  \\       &=&
    \theta(p_0) S_{F}(p) +
    \theta(-p_0) S^{*}_{F}(p) -2 \pi i ~\mbox{sign}(p_0) n_{F}(p) \delta(p^2-m_{D}^2),
\nonumber  \\
    S_{12}(p) &=&  2 \pi i ~\mbox{sign}(p_0)                     n_{F}(p) (\not{\!p} + m_{D}) \delta(p^2-m_{D}^2),
\nonumber  \\
    S_{21}(p) &=& -2 \pi i ~\mbox{sign}(p_0) e^{\beta (p_0-\mu)} n_{F}(p) (\not{\!p} + m_{D}) \delta(p^2-m_{D}^2),
%
\end{eqnarray}
where $S_{F}(p)= (\not{\!p}-m_{D}+i \epsilon)^{-1} $ is the
fermion vacuum Feynman propagator, $n_{F}(p) \equiv (e^{\beta
(p_0-\mu)} + 1)^{-1}$ is the Fermi distribution function, 
and $ J^{\dag} = (J^{\dag}_1, -J^{\dag}_2) $. 
$S$ can be diagonalized by multiplying matrices
$u_{F}$ from both sides
 as $S = u_{F} S_{d} u_{F}^{-1} \eta$, with
\begin{eqnarray}
 S_{d} =
        \left( \begin{array}{cc}
        S_{R}         &   0 \\
        0         &   S_{A}
         \end{array}      \right),
\end{eqnarray}
where $S_R$ and $S_A$ are retarded and advanced Dirac propagators.
Here $u_{F}(p)$ is the thermal Bogoliubov matrix which has the
following form
\begin{eqnarray}
\label{FermionNLuF}
 u_{F}(p)
   =   \sqrt{ n_{F}(p)  } e^{\beta (p_0-\mu)/2}
        \left( \begin{array}{cc}
        1    &    -  e^{-\beta (p_0-\mu)} \\
        1    &    1
               \end{array}      \right).
\end{eqnarray}
Using the above property, one can write
\begin{eqnarray}
     \label{FermionpsiSpsi}
 \psi^{\dag} S^{-1} \psi
 &=& \psi^{\dag} \eta u_{F} S_{d}^{-1} u_{F}^{-1} \psi
 \nonumber  \\ &=&  \bar{\psi} u_{F} S_{d}^{-1} u_{F}^{-1} \psi,
\end{eqnarray}
where $ \psi = (\psi_1, \psi_2) $ is the thermal fermion doublet
 and we define its conjugate
 $ \bar{\psi} \equiv (\psi^{\dag}_1,\psi^{\dag}_2 )\eta$.
By changing the field variables by global Bogoliubov
transformations $u_{F}^{-1} \psi \rightarrow \psi$ and $
\bar{\psi} u_{F} \rightarrow \bar{\psi}$, one can write the
partition function in Eq. 
(A4) without the source
term as
\begin{eqnarray}
     \label{FermionCTP2}
Z
   &=&
    \int d\psi_{f} d\bar{\psi_{f}}
    D\psi D\bar{\psi}
     \exp \left[
       i S_{0}[\psi,\bar{\psi} ]
          \right]
    \rho_{D}[\psi,\bar{\psi}],
\end{eqnarray}
where
\begin{eqnarray}
\label{FermionNLActionBar}
 S_{0}[\psi,\bar{\psi}] &=&
 \bar{\psi} S_{d}^{-1} \psi
 \nonumber  \\ &=&
 \int d^{3}x \frac{d \omega}{2 \pi}
    \left[
 \bar{\psi} (x, \omega)
        \left(
      \omega \gamma^0 -i \vec{\partial} \cdot \vec{\gamma} - m_{D}  + i \epsilon \eta \gamma_0
            \right)
 \psi(x, \omega)
     \right].
\end{eqnarray}
%

%
%
%
%
%
%
%
%
%
%

%
%
In the presence of the random variable $v_{D}$ which is assumed to
obey the probability
 distribution given in Eq. 
 (38), we average the
partition function over $v_{D}$
 and
obtain the reduced action
\begin{eqnarray}
 \langle Z[v_{D}] \rangle
 &=& \int Dv_{D} P[v_{D}] D\bar{\psi} D\psi
  \exp \left[ i S_{v_{D}}[\psi,\bar{\psi}] \right] \nonumber  \\
 &=& \int D\bar{\psi} D\psi
  \exp \left[ i S_{0}[\psi,\bar{\psi}] + i S_I[\psi,\bar{\psi}] \right],
\label{NLSFreducedEA}
\end{eqnarray}
where
\begin{eqnarray}
 S_I[\psi,\bar{\psi}] &=&
   \frac{i}{2} \int d^{4}x d^{4}y
 \bar{\psi}(x) \psi(x)
 \Delta_{D}(x-y)
 \bar{\psi}(y) \psi(y).
\label{NLSFEA}
\end{eqnarray}

By introducing the auxiliary bilocal matrix field
$\sigma_{D}(x,y)$ as
\begin{eqnarray}
\label{NLSFHS}
       e^{i S_I[\psi,\bar{\psi}]  }
   &=&  \\
    \int D \sigma_{D}&
\exp&[\frac{1}{2} \int d^4 x d^4 y \mbox{Tr}[ \sigma_{D}(x,y)
\Delta_{D}^{-1}(x-y) \sigma_{D}(y,x)  ] ] \exp \left[i
S_{HS}[\sigma_{D},\psi,\bar{\psi}]  \right], \nonumber
\end{eqnarray}
where
\begin{eqnarray}
S_{HS}[\sigma_{D},\psi,\bar{\psi}] = - \int d^4 x d^4 y
~\bar{\psi}(x) \sigma_{D}(x,y) \psi(y),
\label{NLSFpart}
\end{eqnarray}
the partition function can be written as
\begin{eqnarray}
  Z
   =  \int D \sigma_{D} D\bar{\psi} D\psi
   \exp \left[ \frac{1}{2} \int \mbox{Tr} [\sigma_{D} \Delta_{D}^{-1} \sigma_{D}]   \right]
   \exp \left[ i S_{0}[\psi,\bar{\psi}] +i S_{HS}[\sigma_{D},\psi,\bar{\psi}]
   \right].
\label{NLSFZpart}
\end{eqnarray}
%
In energy representation, Eqs. (\ref{NLSFpart}) and
(\ref{NLSFZpart}) have the form
\begin{eqnarray}
S_{HS}[\sigma_{D},\psi,\bar{\psi}] =   - \int d^3 x d^3 x' \frac{d
\omega}{2 \pi} \frac{d \omega}{2 \pi}'
~\bar{\psi}(\vec{x}, \omega)
     \sigma_{D \omega \omega'}(\vec{x},\vec{x'})
  \psi(\vec{x'}, \omega')
\label{NLSFsma2}
\end{eqnarray}
and
\begin{eqnarray}
\label{NLSFActionHS2}
 S_{0}[\psi,\bar{\psi}] + S_{HS}[\sigma_{D},\psi,\bar{\psi}]
 &=& \\
\int d^3 x d^3 x' \frac{d \omega}{2 \pi} \frac{d \omega}{2 \pi}'
~\bar{\psi}(\vec{x}, \omega)
 \left[
  \left(
     \omega \gamma^0
   \right.
   \right.
     &-&
   \left.
   \left.
   i \vec{\partial} \cdot \vec{\gamma} - m_{D}  + i \epsilon \eta \gamma_0
  \right)
 \delta(\vec{x} - \vec{x'}) \delta(\omega - \omega')
 - \sigma_{D \omega \omega'}(\vec{x},\vec{x'})
 \right]
\psi(\vec{x'}, \omega').
 \nonumber
\end{eqnarray}

After integrating out $\bar{\psi}$ and $\psi$, we obtain
\begin{eqnarray} \label{PartDiracS}
  Z
   &=&  \int D\sigma_{D}
     \exp \left[ -\frac{1}{2} \int \mbox{Tr} [\sigma_{D} \Delta_{D}^{-1} \sigma_{D}] \right]
\\
  \times &\exp& \left[
  \int d^3 x d^3 x' \frac{d \omega}{2 \pi} \frac{d \omega}{2 \pi}'
 \mbox{Tr} \log
 \left[
  \left(
  \omega \gamma^0 -i \vec{\partial} \cdot \vec{\gamma} - m_{D}  + i \epsilon \eta \gamma_0
  \right)
 \delta(\vec{x} - \vec{x'}) \delta(\omega - \omega')
 - \sigma_{D \omega \omega'}(\vec{x},\vec{x'})
 \right]
               \right]
     \nonumber
\end{eqnarray}
%
%
%
%
%
and, for the time independent stochastic field as in Eq.
(\ref{sigma}),
\begin{eqnarray}
\label{DiracZsigma}
  Z
   &=&  \int D\sigma_{D}
   \exp \left[ -\frac{1}{2u_{D}} \int d^3 x \mbox{Tr} \sigma_{D}^2(x) \right]
\\
  &\times& \exp \left[ \int d^3 x \mbox{Tr} \log
   \{
     \left(
  \omega \gamma^0 -i \vec{\partial} \cdot \vec{\gamma} - m_{D}  + i \epsilon \eta \gamma_0
   \right) \delta(\omega - \omega')
 - \sigma_{D \omega \omega'}(x)
    \}  \right] .\nonumber
\end{eqnarray}
%
%
%
%
%
%
The equation of motion can be obtained from Eq.
(\ref{DiracZsigma}) by functional derivative with respect to
$\sigma_{D}$. Following the steps from Eq. (\ref{NLSBSSterm}) to
Eq. (\ref{NLSBssmomentum2}), the kinetic term in the $\sigma_{D}$
field in Eq. (\ref{PartDiracS}) can be given similarly to Eq.
(\ref{NLSBssmomentum2}) which enables us to constuct the effective
field theory in terms of the collective field.


By the Kubo formula the conductivity can be written in terms of
the Green functions as in Eq. (\ref{NLSBnoncondG})
\begin{eqnarray}\label{noncondS}
C_{\kappa}(\vec{x},\vec{y}) &  \equiv & \\
 - \frac{1}{4 \pi \kappa}
\int_{0}^{\infty} d \omega \Omega_{\kappa}(\omega)& \mbox{Tr} & \{
\left[ S_R(\vec{x},\vec{y},\omega) -  S_A(\vec{x},\vec{y},\omega)
\right] \vec{\gamma}
 \left[ S_R(\vec{y},\vec{x},\omega+\kappa)
     -  S_A(\vec{y},\vec{x},\omega+\kappa) \right]
\vec{\gamma} \}. \nonumber
\end{eqnarray}
We obtain
\begin{eqnarray} \label{noncondpsi}
C_{\kappa}(\vec{x},\vec{y})& &\equiv \nonumber  \\
  \frac{-1}{4 \pi \kappa} \int_{0}^{\infty} d\omega 
  \Omega_{\kappa}(\omega)& &
\sum_{abcd}  \langle \psi^{\dag a}(\vec{x},\omega) \vec{\gamma}
\psi^{b}(\vec{x},\omega + \kappa) \psi^{\dag c}(\vec{y},\omega +
\kappa) \vec{\gamma} \psi^{d}(\vec{y},\omega) \rangle.
\end{eqnarray}

 This expression
can be also obtained directly from the partition function by
functional derivative after introducing the source term in the
form
\begin{equation}\label{source}
 i \int d^3 x \frac{d \omega}{2 \pi} \frac{d \omega}{2 \pi}'
~\bar{\psi}(\vec{x}, \omega) \vec{A}^{\kappa} \cdot \vec{\gamma}
\delta_{\kappa}
  \psi(\vec{x}, \omega'),
\end{equation}
where $\vec{A}^{\kappa}$ is the external source field and
$\delta_{\kappa}$ was defined in Eq. (\ref{Lmatrix}). Then the
conductivity is given similarly by Eq. (\ref{NLSBcondderiv}). The
gauge transformation
\begin{eqnarray}
      \psi(\vec{x})  &\rightarrow&
      e^{-i \vec{x} \cdot \vec{A}^{\kappa} \delta_{\kappa}}  \psi(\vec{x}), \nonumber  \\
 \bar{\psi}(\vec{x}) &\rightarrow& \bar{\psi}(\vec{x})
      e^{i \vec{x} \cdot \vec{A}^{\kappa} \delta_{\kappa}},
      \label{Diracgaugecoupl}
\end{eqnarray}
as we saw in Eq. (\ref{NLSBgaugecoupl}),
generates the gauge coupling in the effective theory represented by
\begin{eqnarray} \label{DiracZsigma2}
  Z &=& \int D\sigma_{D}
     \exp \left[-\frac{1}{2} \int \mbox{Tr} [\sigma_{D} \Delta^{-1} \sigma_{D}] \right]
     \exp \left[
  \int d^3 x d^3 y \frac{d \omega}{2 \pi} \frac{d \omega}{2 \pi}'
    \right.
 \\  \mbox{Tr}& \log & \left.
   \{
   \left[
   \left(
   \omega \gamma^0 -i \vec{\partial} \cdot \vec{\gamma} - m_{D}
   + i \epsilon \eta \gamma_0
   \right) \delta(\omega - \omega')
  + i \vec{A}^{\kappa} \cdot \vec{\gamma} \delta_{\kappa}
  \right] \delta(\vec{x} - \vec{y})
 -\sigma_{D \omega \omega'}(\vec{x},\vec{y})
   \}
   \right]. \nonumber
\end{eqnarray}
This allows us to write the mesoscopic fluctuations in terms of
$\sigma_{D}$ fields. We can discuss universal fluctuations
parallel to the scalar fields in this formalism. For an initially
thermal equilibrium state, $\Omega_{\kappa}(\omega) =
\rho_{D}[\omega]-\rho_{D}[\omega+\kappa]$, where $\rho_{D}[\omega]
= n_F(\omega)$ is the Fermi distribution, the dc limit $\kappa
\rightarrow 0$ extracts the Fermi energy $\omega_F$ that reminds
us of the electron transport problem.

%
%

\section{Conductance fluctuations}
\label{app:Cf}
\begin{eqnarray}\label{AppC1}
 C^{(1)} &=&
\frac{2 \pi^4}{\kappa_{1} \kappa_{2} } \frac{1}{(2 \pi)^{5}}
 \int d \omega d \omega' \Omega_{\kappa}(\omega-\kappa) \Omega_{\kappa}(\omega')
N^2(k_0) D^2(k_0) \int d^3 P \\
 &\left[
\right.& \left. \langle
 q_{\omega-\kappa_1~\omega'}(\vec{P})
 q^{\dag}_{\omega'~\omega-\kappa_1}(\vec{-P})
\rangle + h.c. \right]
\left[
\langle
 q_{\omega'+\kappa_2~\omega}(\vec{P})
 q^{\dag}_{\omega~\omega'+\kappa_2}(\vec{-P})
\rangle + h.c. \right]. \nonumber
%
\end{eqnarray}
For $\kappa_{1} \kappa_{2} \rightarrow 0$,
\begin{eqnarray}\label{AppC12}
 C^{(1)} &\rightarrow&
\frac{2^5}{(2 \pi)^2} \int_{M_{IR}}^{\infty} \frac{d p}{p^2} =
\frac{8}{\pi^2 M_{IR}}.
\end{eqnarray}
\begin{eqnarray}\label{AppC2}
 C^{(2)} &=&
  \frac{4 \pi^5}{\kappa_{1} \kappa_{2}}
  \frac{1}{(2 \pi)^{5}}
  \int d \omega d \omega' \Omega_{\kappa}(\omega-\kappa) \Omega_{\kappa}(\omega')
N^3(k_0) D^3(k_0) \int d^3 P  \vec{P}^2 \\
&\left[ \right. & \left.
\langle
 q_{\omega-\kappa_1~\omega'}(\vec{P})
 q^{\dag}_{\omega'~\omega-\kappa_1}(\vec{-P})
\rangle
\langle
 q_{\omega'+\kappa_2~\omega}(\vec{-P})
 q^{\dag}_{\omega~\omega'+\kappa_2}(\vec{P})
\rangle \langle
 q_{\omega-\kappa_2~\omega'+\kappa_1}(\vec{P})
 q^{\dag}_{\omega'+\kappa_1~\omega-\kappa_2}(\vec{-P})
\rangle \right. \nonumber  \\ &+& \left.  h. c. \right]. \nonumber
\end{eqnarray}
For $\kappa_{1} \kappa_{2} \rightarrow 0$,
\begin{eqnarray}\label{AppC22}
 C^{(2)} &\rightarrow&
\frac{2^7}{(2 \pi)^2} \int_{M_{IR}}^{\infty} \frac{d p}{p^2} =
\frac{32}{\pi^2 M_{IR}}.
\end{eqnarray}
\begin{eqnarray}\label{AppC3}
C^{(3)} &=&
  \frac{\pi^6}{\kappa_{1} \kappa_{2}}
  \frac{1}{(2 \pi)^{5}}
  \int d \omega d \omega'
  \Omega_{\kappa}(\omega-\kappa) \Omega_{\kappa}(\omega')
N^4(k_0) D^4(k_0) \int d^3 P  \vec{P}^4 \\
 &\left[ \right.& \left.
\langle
 q_{\omega-\kappa_1~\omega'}(\vec{P})
 q^{\dag}_{\omega'~\omega-\kappa_1}(\vec{-P})
\rangle
\langle
 q_{\omega'+\kappa_1~\omega}(\vec{-P})
 q^{\dag}_{\omega~\omega'+\kappa_1}(\vec{P})
\rangle \langle
 q_{\omega' \omega}(\vec{-P})
 q^{\dag}_{\omega \omega'}(\vec{P})
\rangle  \right.
\nonumber  \\
&\times& \left.
\langle
 q_{\omega-\kappa_1~\omega'+\kappa_2}(\vec{P})
 q^{\dag}_{\omega'+\kappa_2~\omega-\kappa_1}(\vec{-P})
\rangle + h. c. \right] + \mbox{three other terms}. \nonumber
\end{eqnarray}
For $\kappa_{1} \kappa_{2} \rightarrow 0$,
\begin{eqnarray}\label{AppC32}
C^{(3)} &\rightarrow& \frac{2^5 \times 9}{(2 \pi)^2}
\int_{M_{IR}}^{\infty} \frac{d p}{p^2} = \frac{72}{\pi^2 M_{IR}}.
\end{eqnarray}

\end{document}